# The Jalisco Seismic Accelerometric Telemetric Network (RESAJ)

by Francisco Javier Núñez-Cornú, Juan Manuel Sandoval, Edgar Alarcón, Adán Gómez, Carlos-Plascencia Suarez, Diana Núñez Escribano, Elizabeth Trejo-Gómez, Oscar Sánchez Mariscal, J. Guadalupe Candelas Ortiz, and Luz María Zúñiga-Medina

## ABSTRACT

The Jalisco region of western Mexico is the locus of interaction among the North America, Cocos, and Rivera plates, giving rise to the Jalisco block. This region is one of the most tectonically active in Mexico, and here took place the largest instrumentally recorded earthquake in Mexico the twentieth century, on 3 June 1932 (M 8.2), three important tsunamis in the last 100 yrs, and two of the most active volcanoes in Mexico. Nevertheless, the first seismicity studies here, undertaken with temporary networks, did not commence until 1994. In 2008, the Government of Jalisco and the University of Guadalajara funded a research project to install a seismic network in this region. The principal objective was to study the seismic hazard in the region and characterize seismic parameters in the different areas to design building codes. The Red Sísmica y Acelerométrica Telemétrica de Jalisco (RESAJ) project was thus initiated in 2009. Its Central Lab is at Centro de Sismología y Volcanología de Occidente (SisVOc), located at the Universidad de Guadalajara in Puerto Vallarta. Currently, the RESAJ has 26 telemetered and 2 autonomous stations. The RESAJ serves as the seismological lab for the postgraduate program at SisVOc.

## INTRODUCTION

The Jalisco region is vulnerable to different natural hazards, among them destructive earthquakes that pose particular risk because the region hosts Mexico's second largest city, the Guadalajara metropolitan zone, which has very important technological and industrial infrastructure. The only way to mitigate the potentially disastrous effects from seismic, volcanic, and tsunami hazards in such areas is to carry out scientific investigation to better understand the seismic sources and their likely impacts. Accordingly, the study of geological and tectonic causes—structural, kinematic, and dynamic characteristics, and potential destructive effects—is indispensable.

The Jalisco block (JB) region (Figs. 1 and 2), characterized by high seismicity, has experienced numerous destructive earthquakes of great magnitude. The largest instrumentally recorded historic earthquake occurred in Mexico during the twentieth century, on the coast of Jalisco in 1932 (number 15 in Table 1 and Fig. 3). This event was followed by a magnitude 7.8 earthquake (number 16 in Table 1 and Fig. 3) 15 days later. Singh et al. (1985) propose that the rupture area for both earthquakes was the entire coast of Jalisco, from Bahía de Banderas to Colima graben. In 1995, an earthquake of magnitude 8.0 occurred off the coast of Jalisco, but this earthquake (number 21 in Table 1 and Fig. 3) ruptured only the southern half of the area proposed for the 1932 events (Fig. 1), suggesting that the northern coast of Jalisco, including Bahia de Banderas, includes a seismic gap which may be poised to fail (Fig. 1). There have also been large intraplate earthquakes in the region, such as the events of 30 December 1567 (number 1 in Table 1 and Fig. 3) and 11 February 1875 (number 10 in Table 1 and Fig. 3). Additional hazards in the region include three important tsunamis in the last 100 yrs (Trejo-Gómez et al., 2015) and three active volcanoes: Sanganguey, Ceboruco, and the most active volcano historically in Mexico, Volcan de Fuego (also known as Colima or Zapotlán), as well as the volcanic structure known as Caldera de la Primavera (Fig. 2).

The main objective of the Red Sísmica y Acelerométrica de Jalisco (RESAJ) project is to enhance our capability to research and study the potential for destructive earthquakes within the Jalisco region by deploying a modern seismic network. This network allows us to study the Rivera plate (RP) and JB seismicity, including active volcanoes and seismic tsunamigenic sources.

## TECTONIC SETTING

According to Núñez-Cornú et al. (2002), the North America, Pacific, Cocos, and Rivera lithospheric plates interact in the western Mexican volcanic belt. Several triple-junction locations have been proposed (Fig. 1), but the seismotectonic processes at work are still not fully understood. The existence of a tectonic unit in this region, known as the JB, has been proposed by several researchers (Luhr et al., 1985; Bourgois et al., 1988; DeMets and Stein, 1990; Allan et al., 1991; Garduño and

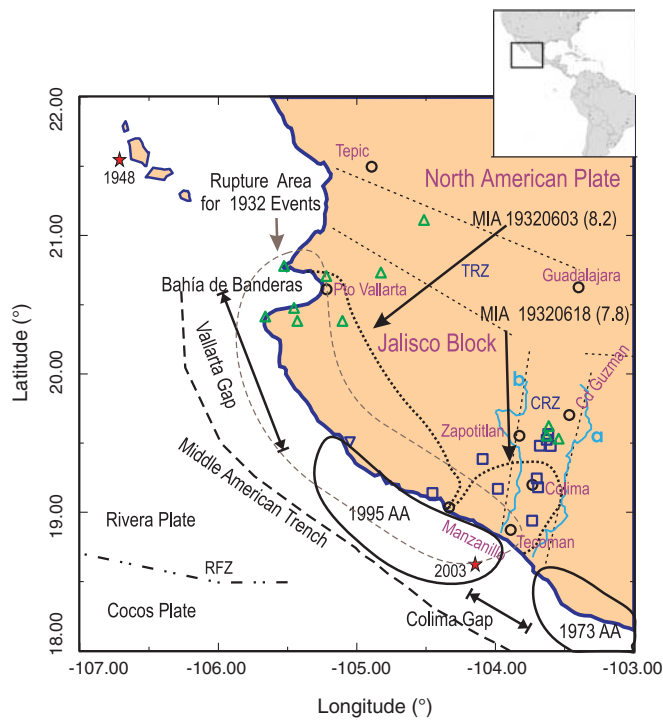

▲ **Figure 1.** Seismotectonic framework along the Jalisco coast (modified from Trejo-Gómez et al., 2015). RFZ, Rivera fault zone; CRZ, Colima rift zone; TRZ, Tepic–Zacoalco rift zone; a, Armería River; b, Cohuayana River; MIA, maximum intensity areas for earthquakes in 1932 (dates and magnitudes indicated); and AA, aftershocks areas. Circles indicate locations of cities, and stars show the epicenters of the 4 December 1948 and the 2003 Armería earthquakes; seismic gaps proposed; green triangles indicate Red Sismológica de Jalisco (RESJAL) stations; blue squares indicate Colima Seismic Telemetric Network (RESCO) stations; inverted blue triangle indicates Chamela Servicio Sismologico Nacional (SSN) station.

Tibaldi, 1991). The JB is limited to the east by the Colima rift zone, which extends northward from the Pacific coast and connects at its northern end with two other major extensional structures: the Tepic-Zacoalco rift zone (TRZ; trending roughly northwest–southeast), defined as the northern boundary of the JB, and the Chapala rift zone (trending roughly east–west). The connection between the northwestern border of the JB and the continent (the Tamayo fault system) is not well defined. This border has been related to the San Blas fault as continuation of the TRZ, or to the Islas Marias Escarpment (IME), west of Tres Marias Islands (Fig. 2). Recent studies (Dañobeitia et al., 2016; Núñez-Cornú et al., 2016) indicate that in the north of Tres Marias Islands (from north to south: Maria Madre, Maria Magdalena, and Maria Cleofas) there is no clear evidence of an active subduction zone; instead, faulting at the west flank of the Tres Marias Islands is observed, while southward, between Maria Magdalena and Maria Cleofas Islands, the subducted slab of the RP is clearly delineated by the seismicity.

Núñez-Cornú et al. (2016) show the existence of a tectonic feature south of Maria Cleofas island, the Sierra de Cleofas (SC). This 100-km-long structure is oriented north–south and marks the limit between the RP and the JB, possibly arising from compression by the RP against the JB. It suggests the beginning of present-day subduction, with associated seismic activity. These authors also suggest that Banderas Canyon (BC) is apparently in a reflection of extension from west to east that seems to continue through the Rio Pitillal river valley. There is no seismic or morphological submarine evidence to indicate that BC is a continuation of the Valle de Banderas or Vallarta graben. The shape of BC is controlled by continental stresses. Mortera Gutiérrez et al. (2016) and Núñez-Cornú et al. (2016) show that the Bahía de Banderas area is undergoing strong crustal stress via convergence of the RP (Kostoglodov and Bandy, 1995). The existence of shallow submarine hydrothermal activity in the Bahía de Banderas (Núñez-Cornú et al., 2000) could be a result of these stresses. Urías Espinosa et al. (2016) suggest that the existence of Ipala Canyon (IC) is related to extension produced by the sharp change in the RP convergence, and IC may lay along the southeast boundary of a major fore-arc block, termed the Banderas fore-arc block (BFB; Fig. 2).

## BACKGROUND

Historic macroseismic data for the Jalisco region date back to 1544 (Núñez-Cornú, 2011). In the past 450 yrs (Table 1; Fig. 3), there have been at least 22 major earthquakes with $M > 7.0$ (Núñez-Cornú, 2017). The recurrence times for earthquakes, such as the 1932 event on the coast of Jalisco, have been estimated by Singh et al. (1985) as 77 yrs (±15%). As mentioned above, the earthquake of 1995 ruptured only the southern half of the 1932 rupture, which implies that the probability of an equivalent event filling the northern gap may be high (Fig. 1). On 22 January 2003, there was a shallow earthquake of $M_w$ 7.4 on the continental shelf (Fig. 1), not associated with subduction (Núñez-Cornú et al., 2004, 2010). The tectonic complexity of this region, including possible oblique subduction of the RP, provides a strong likelihood for the existence of still unrecognized tectonic structures capable of generating earthquakes of moderate-to-large magnitude (7.0–7.6); such an unanticipated source produced an earthquake of M ∼ 7.0 near the Islas Marias (number 18 in Table 1 and Fig. 3). This earthquake resulted in massive destruction at Maria Madre Island.

Despite the risk associated with these tectonic processes, until 2001 routine seismic monitoring for the area was limited to just one permanent seismic station from Servicio Sismologico Nacional (SSN) at Chamela, on the coast of Jalisco, and the Colima Seismic Telemetric Network (RESCO) of the University of Colima (UCol), located on Colima Volcano and the southern part of the Colima graben (Fig. 1).

In 1994, local seismicity studies initiated using a portable seismic network in a joint project between Centro de Investigación Científica y Educación Superior de Ensenada

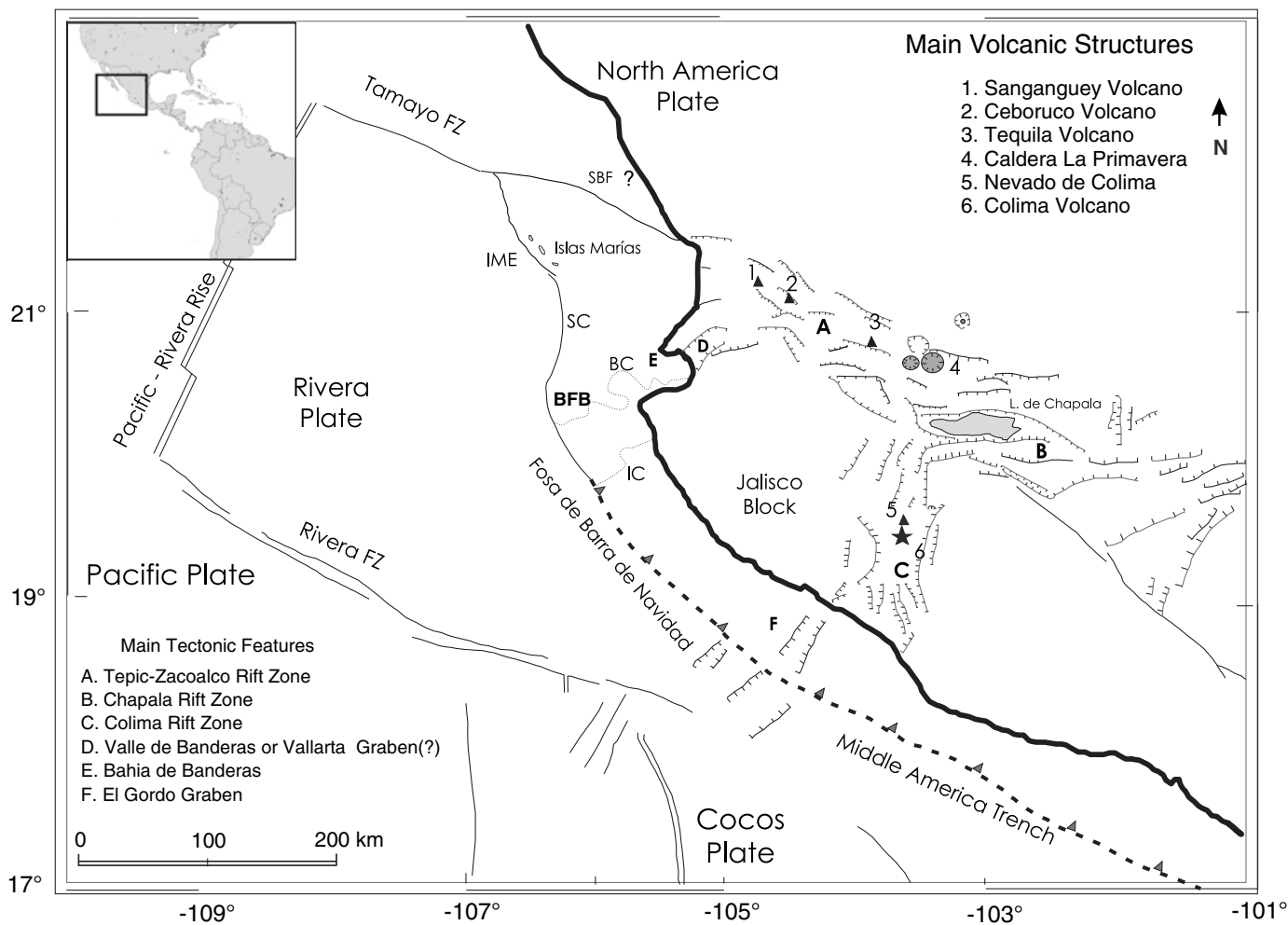

▲ Figure 2. Tectonic framework of western Mexico: Rivera and Cocos plates subducting beneath the North American plate and transform faults zones associated with spreading. SBF, San Blas fault (proposed); IME, Islas Marías Escarpment; SC, Sierra de Cleofas; BC, Banderas Canyon; IC, Ipala Canyon; BFB, Banderas fore-arc block.

(CICESE) and the Universidad de Guadalajara (UdeG), funded by Consejo Nacional de Ciencia y Tecnología (CONACyT) with the support of Jalisco Civil Defense (PCJal; Núñez-Cornú et al., 2002). These authors showed for the first time the high level of seismicity within the JB. On this basis, the UdeG and PCJal began to deploy the Red Sismológica de Jalisco (RESJAL) in late 2001. The RESJAL included 10 digital 3D seismograph stations, distributed between the states of Jalisco and Nayarit, of which six were telemetered and four were autonomous (Fig. 1). The network included sites on Colima and Ceboruco volcanoes. RESJAL operated only until late 2004.

Using data from RESJAL and RESCO, Rutz and Núñez-Cornú (2004) showed the high level of seismicity in the region and found, at that time, that the SSN only reported earthquakes with magnitude greater than $M_L$ 4.25, and failed to report 45% of these. Most of the seismicity reported by Rutz and Núñez-Cornú (2004) show magnitudes smaller than 4.0. Differences in epicentral locations reported by Rutz and Núñez-Cornú (2004) and those reported by the SSN may be as great as 80 km in some parts of the region. Differences in hypocentral depths for inland earthquakes average 40 km. Rutz and Núñez-Cornú (2004) concluded that, at that time, the data of the SSN were of insufficient quality for accurate seismotectonic models, and they recognized the necessity to deploy a permanent local seismic network to better assess the seismic hazard in the region.

Project "Mapping the Riviera Subduction Zone" was conducted from January 2006 to June 2007 in a collaboration between American and Mexican institutions. This project deployed a temporary seismic network using 50 broadband stations within the states of Michoacán, Colima, and Jalisco. Data obtained in this project have been used in a variety of studies (Gutierrez et al., 2015; Ochoa-Chávez et al., 2016; Pinzon et al., 2017).

In 2009 PCJal, collaborating with UCol, installed a telemetered seismic network in Jalisco that included a tsunami warning system with support from the Centro Nacional de Prevencion de Desastres, Fondo Nacional de Prevención de Desastres, and the state of Jalisco (PCJal, 2016).

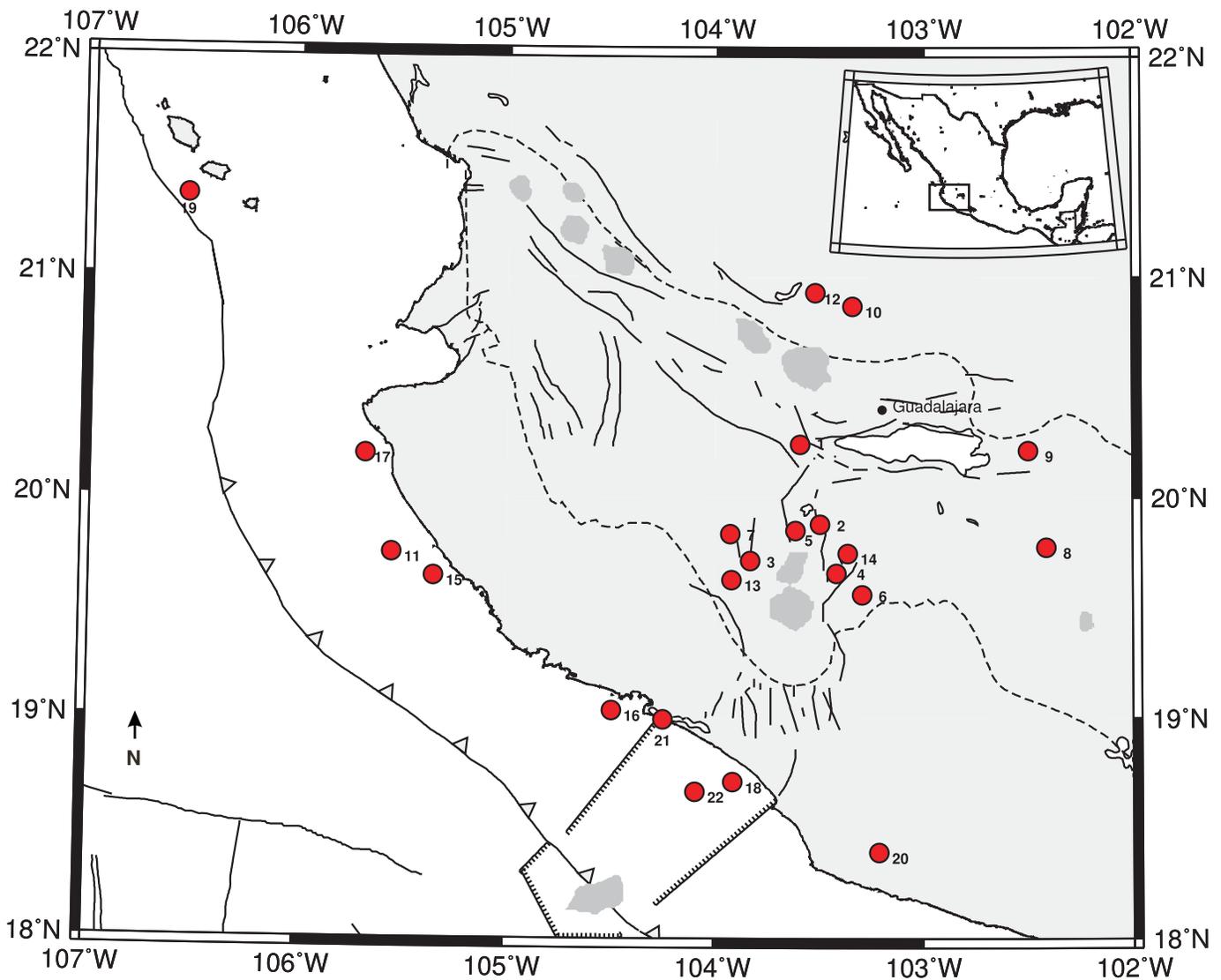

▲ Figure 3. Historical earthquakes near or within the Jalisco block (Table 1).

The government of the state of Jalisco approved funds in 2008 to develop a seismic network to study the seismic hazard associated with tectonic processes in Jalisco to better assess the regional seismic risk and aid in the design of adequate building codes. The Consejo Estatal de Ciencia y Tecnología de Jalisco and the UdeG supported this effort as a research project. By mid-2009, the Jalisco Seismic Accelerometric Telemetric Network (RESAJ) was thus initiated.

The criteria for the selection of the sites in the deployment of the stations were
- geographic distribution according to seismicity pattern reports in previous works (Núñez-Cornú *et al.*, 2002, 2004),
- access to telemetry (line of view) or internet,
- security of the instruments, and
- seismic noise.

Each RESAJ station has a 24 bit A/D, 6 channels Quanterra Q330-6ch or Q330S-6ch DAS digitizer; a Lennartz LE3D 1Hz seismometer (Lennartz Electronic GmbH, 2016); and a Kinemetrics triaxial accelerometer episensor Model FBA ES-T and solar power supply (Fig. 4). Eleven of the seismic stations will have a Global Positioning System Trimble NetR9 Global Navigation Satellite System (GNSS) reference receiver. The data are transmitted to the Central Lab at Centro de Sismología y Volcanología de Occidente (SisVOc) in Puerto Vallarta (PV) using freewave type ethernet radios, wireless LAN links (Mikrotic type) and public internet links. A typical shed housing the seismic stations is shown in Figure 5. Data are handled by the Antelope System software (Lindquist *et al.*, 2007); sample rate for seismometers is 100 Hz and sample rate for accelerometer is 200 Hz.

RESAJ deployment was planned in three stages as follows:
1. deploy 10 seismic stations and install the Central Lab, where data are received, processed, and stored at SisVOc in PV;
2. deploy 15 additional seismic stations and tune up the network; and

## Table 1
### Historic Earthquakes in the Jalisco Region

| Number | Date (yyyy/mm/dd) | Time (hh:mm) | Latitude (°) | Longitude (°) | Magnitude | References |
|---|---|---|---|---|---|---|
| 1 | 1567/12/30 | | 20.2423 | −103.5937 | 7.2 | Suter (2015) |
| 2 | 1577/12/28 | | 19.8793 | −103.4964 | 7.0 | Núñez-Cornú (2017) |
| 3 | 1587/02/20 | | 19.7138 | −103.8315 | 7.0 | Núñez-Cornú (2017) |
| 4 | 1611/04/15 | | 19.6576 | −103.4180 | 7.0 | Núñez-Cornú (2017) |
| 5 | 1749/10/22 | 16:00 | 19.8511 | −103.6141 | 7.0 | Núñez-Cornú (2017) |
| 6 | 1806/03/26 | 16:15 | 19.5600 | −103.2951 | 7.5 | Núñez-Cornú (2017) |
| 7 | 1818/05/31 | 03:07 | 19.8358 | −103.9273 | 7.7 | Núñez-Cornú (2017) |
| 8 | 1837/11/22 | 23:50 | 19.7752 | −102.4116 | 7.7 | Núñez-Cornú (2017) |
| 9 | 1847/10/02 | | 20.2143 | −102.4988 | 7.1 | Núñez-Cornú (2017) |
| 10 | 1875/02/11 | 20:30 | 20.8688 | −103.3425 | 7.8 | Núñez-Cornú (2017) |
| 11 | 1875/03/09 | 09:21 | 19.7479 | −105.5487 | 8.2 | Núñez-Cornú (2017) |
| 12 | 1878/03/22 | 07:25 | 20.9288 | −103.5219 | 7.1 | Núñez-Cornú (2017) |
| 13 | 1900/01/20 | 06:33 | 19.6270 | −103.9191 | 7.8 | Núñez-Cornú (2017) |
| 14 | 1911/06/07 | 04:26 | 19.7471 | −103.3627 | 7.9 | Singh et al. (1984) |
| 15 | 1932/06/03 | 10:36 | 19.6420 | −105.3477 | 8.2 | Núñez-Cornú (2017) |
| 16 | 1932/06/18 | 10:12 | 19.0340 | −104.4925 | 7.8 | Núñez-Cornú (2017) |
| 17 | 1934/11/30 | 02:05 | 20.1949 | −105.6791 | 7.2 | Núñez-Cornú (2017) |
| 18 | 1941/04/15 | 19:09 | 18.7098 | −103.9114 | 7.9 | Núñez-Cornú (2017) |
| 19 | 1948/12/03 | 00:22 | 21.1670 | −106.7830 | 7.0 | USGS |
| 20 | 1973/01/30 | 21:01 | 18.3900 | −103.2100 | 7.5 | Singh et al. (1984) |
| 21 | 1995/10/09 | 15:35 | 18.9930 | −104.2450 | 8.0 | SSN |
| 22 | 2003/01/22 | 02:06 | 18.6658 | −104.0895 | 7.8 | Núñez-Cornú et al. (2004) |

Later events were instrumentally recorded and located; events prior to 1948 have approximate locations based on macroseismic information. USGS, U.S. Geological Survey; SSN, Servicio Sismologico Nacional.

3. install local networks at Ceboruco and Colima volcanoes. Install GNSS equipment in some of the seismic stations.

The first stage was completed in October 2011 when the Central Lab began to operate at SisVOc. Figure 6 shows Antelope automatic locations for the first two months of operation (25 October 2011–25 January 2012). The system reported 840 detections; for the same period, the SSN reported 107 events and the International Seismological Centre reported 93 events (Núñez-Cornú et al., 2011).

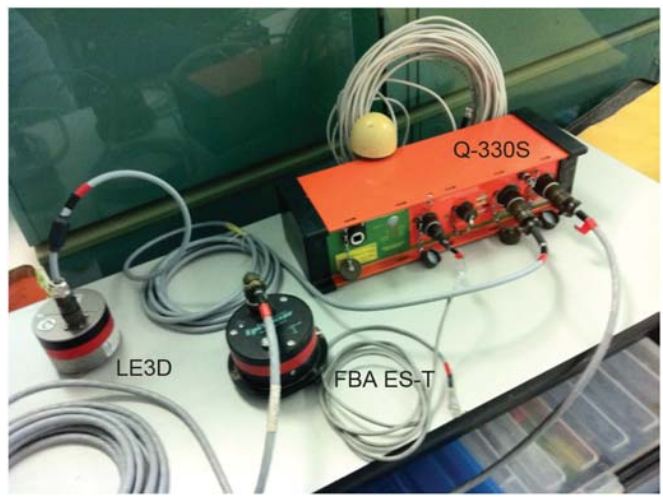

▲ Figure 4. Seismic equipment at each station. Quanterra Q330S datalogger; Lennartz LE3D seismometer, and Episensor FBA ES-T accelerometer.

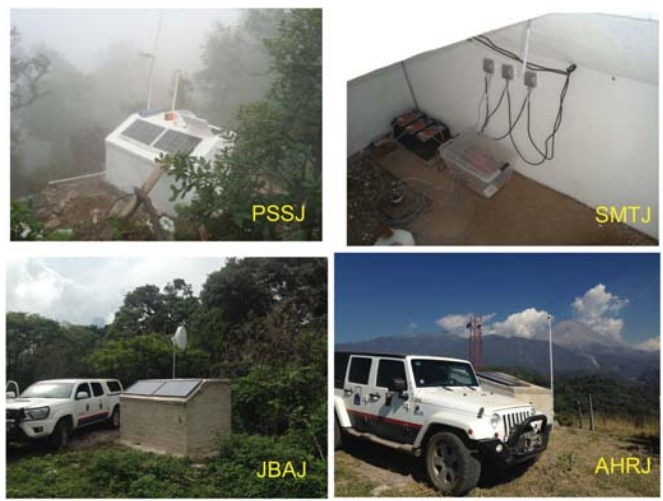

▲ Figure 5. Seismic stations sheds for housing equipment at each site.

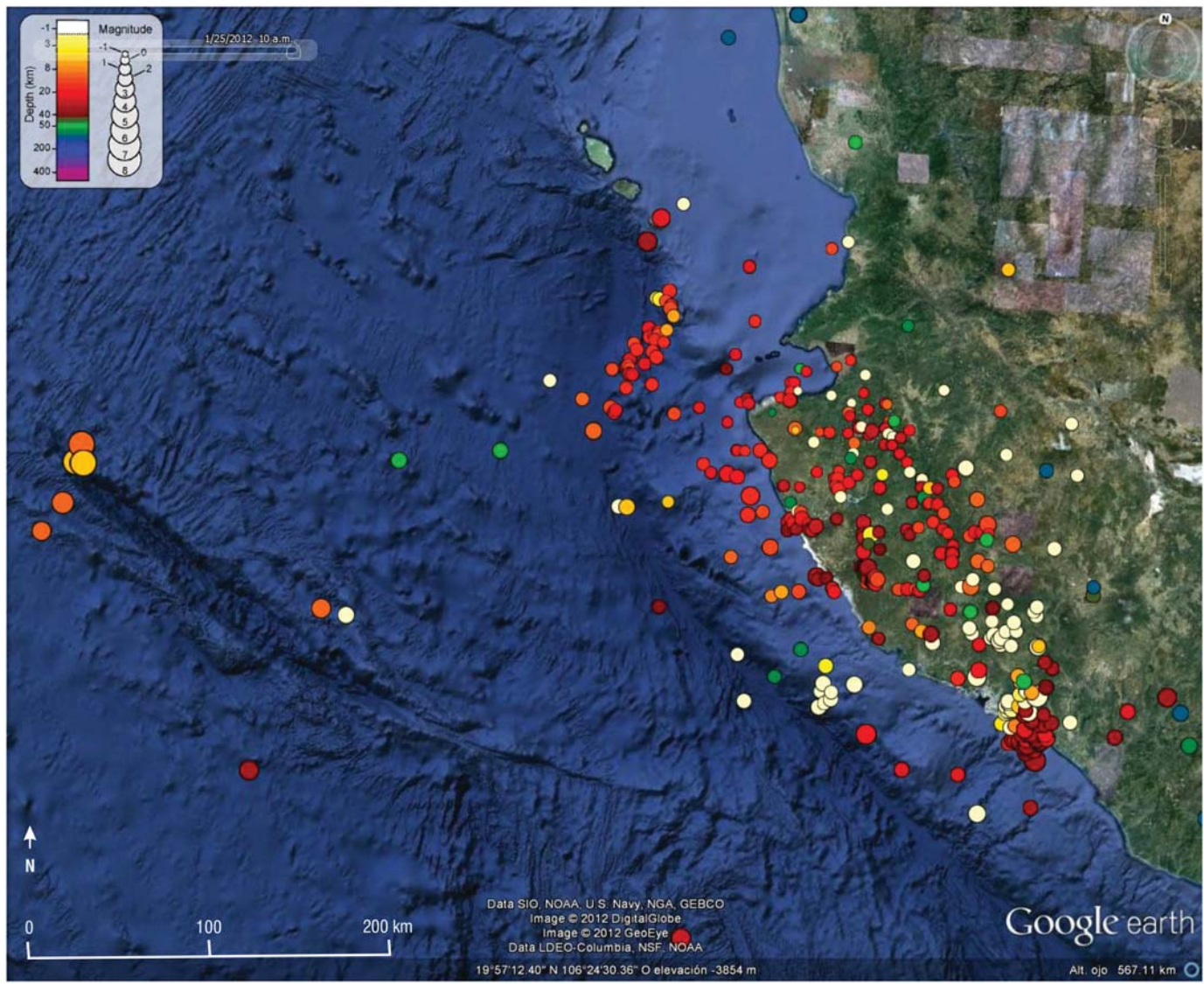

▲ **Figure 6.** Antelope software automatic earthquake short-term average/long-term average locations for period between 25 October 2011 and 26 January 2012.

The greatest challenge for the RESAJ network was achieving real-time data transmission to SisVOc for all stations in a region that has very complex topography. The first plan was to create small network nodes or cells, wherein the stations had direct line of sight to a foreign campus of UdeG in different locations; from these nodes, the data would be sent through the UdeG internal internet network to SisVOc. The poor quality of the internal internet network, however, and differences in policies at the different campuses and the UdeG Central Administration in Guadalajara (CAG), rendered this approach unworkable. Commercial data transmission (satellite, telephone, mobile, private links, etc) is not feasible for a research project, so we use a wireless link network using radio repeaters to communicate to SisVOc. Repeater installation for the telemetry was problematic because most advantageous sites are occupied, and the cost of tower access is prohibitive, as is rental of land for our own tower. The UdeG has since built a tower at La Primavera (PR; Fig. 1) to develop a wireless link between some of its campuses, so we are leveraging that facility to connect our links and send the signal to the CAG and then to PV through the UdeG internal internet network (UGI), which has improved. Some gaps in the transmission still occur, but Antelope Software can recover most of the data using Q330S internal memory (32 MB). For redundancy, most of the stations also possess a dual USB memory (32 GB) that records the data on site, which can be recovered using FTP protocol or visiting the station. We are using PCJal facilities at Nevado de Colima to install a repeater; however, there is a significant lighting problem there that we are currently working to solve. In the second stage of the project, we divide the network into two large cells, one in the north and on the coast, with stations that can be linked directly to PV, and the second cell in the south and center of the region sends the signal to PR and then to CAG.

| Table 2 Locations of RESAJ Stations |||| 
|---|---|---|---|
| Station | Latitude (°) | Longitude (°) | H (m) |
| AHRJ | 19.584 | −103.725 | 1580 |
| ALGJ | 20.388 | −105.479 | 424 |
| AMEJ | 20.507 | −104.005 | 1279 |
| AUTJ | 19.751 | −104.385 | 1194 |
| CAFJ | 19.535 | −103.523 | 2123 |
| CEBJ | 21.113 | −104.508 | 2123 |
| CORJ | 20.414 | −105.661 | 158 |
| CPRJ | 20.595 | −103.545 | 2276 |
| CUCJ | 20.704 | −105.222 | 12 |
| GOFJ | 19.781 | −103.462 | 1652 |
| IMCJ | 21.324 | −106.240 | 542 |
| IMMJ | 21.624 | −106.584 | 606 |
| JAPJ | 19.489 | −103.705 | 1788 |
| JBAJ | 19.489 | −103.545 | 1756 |
| JUAJ | 20.051 | −103.658 | 2898 |
| LLAJ | 19.548 | −103.961 | 2120 |
| MCUJ | 20.382 | −105.104 | 2192 |
| MZCJ | 19.897 | −103.004 | 2865 |
| PHOJ | 19.231 | −103.414 | 844 |
| PSSJ | 20.717 | −104.827 | 2557 |
| PV1J | 20.606 | −105.198 | 544 |
| RESJ | 20.143 | −105.508 | 200 |
| SANJ | 21.465 | −104.767 | 1423 |
| SMTJ | 19.572 | −103.324 | 1298 |
| TAMJ | 19.246 | −104.733 | 120 |
| TEQJ | 20.787 | −103.846 | 2920 |
| TOMJ | 19.970 | −105.253 | 185 |
| YLTJ | 20.477 | −105.466 | 616 |
| RESAJ, Red Sísmica y Acelerométrica Telemétrica de Jalisco. |||| 

In the second stage, we had problems at some stations due to lightning strikes; some of the Q330S with lighting damage could be repaired, others could not. To avoid this problem, we are changing all the electrical grounding systems at all stations. Two complete stations were stolen, and we had to relocate and reinstall one of them (OLOJ). After 10 yrs of operating an autonomous station (PMI) at Punta Mita, Nayarit, at the northwest point of Bahía Banderas, our station had to be removed because the resort decided that a seismic station was not good for tourism.

## PRESENT NETWORK CONFIGURATION AND OPERATIONS

Currently, the first two stages of installation are complete, and we are working on the third. We have 28 stations (Table 2; Figs. 7 and 8), 26 transmitted in real time to SisVOc and 2 autonomous stations (one at Maria Madre Island [IMMJ] and one at Maria Cleofas Island [IMCJ]; Fig. 1). This year (2017) the old autonomous station at Ceboruco volcano (CEBN) was changed to a RESAJ telemetered station (CEBJ) and we installed a RESAJ station at Tequila volcano (TEQJ), facilitating the installation of a telemetric link between both cells and transmitting the data directly to PV. We will continue to maintain the link PR–CAG for redundancy.

We planned to install two additional stations, one between TOMJ and TAMJ and another between MCUJ and AUTJ, but the problem of the telemetry link (Fig. 7) remains unsolved. Deployment of the GNSS stations will be completed in early 2018. The Antelope system controls data acquisition in real time and generates the database. Antelope uses a short-term average/long-term average algorithm for preliminary locations using the IASP91 P-wave velocity model. For relevant earthquakes, P- and S-wave readings are reviewed and corrected by a seismologist, and events are relocated using Hypo71 and local P-wave velocity model (Núñez-Cornú et al., 2002).

The site of the Central Lab is equipped with two Macintosh servers on which the Antelope system software is installed for seismic data acquisition and automatic processing; we maintain one Dell server with an Earthworm system, one DELL server for GNSS data acquisition, and two NAS (24 and 48 TB) for data storage.

Our station distribution allows us to improve our understanding of seismotectonic processes in the following geographic areas (Figs. 1 and 2):

- Bahía de Banderas - Islas Marías, where tsunamigenic hazard is present (Núñez-Cornú et al., 2016);
- the northern coast of Jalisco, which is thought to currently host a seismic gap of concern, and the southern coast;
- the seismicity along the Colima and Tepic–Zacoalco grabens; and
- the intraplate seismicity within the JB.

With the actual stations distribution, the Antelope automatic detection algorithm has an M ≥ 2 level detection in all of the region; in some areas, the level of detection is M ≥ 1.

To monitor volcano seismicity, we have a RESAJ station at Sangangüey Volcano, Ceboruco Volcano, and Caldera La Primavera. Additionally, we installed three additional telemetered stations (Obsidian Kinemetrics dataloggers with LE3D 1 Hz sensors) at Ceboruco Volcano, all within a radius of 8 km from CEBJ, to provide a small network on the volcano. On Volcán de Fuego (Colima), there are four stations (Q330S 3ch with LE3D 1 Hz sensor) and five RESAJ stations nearby. This distribution of stations allows us to identify any change in the seismic pattern at the three volcanoes, including detection and tracking of distal earthquakes, which recent studies have suggested may be a harbinger of new eruptive activity (White and McCausland, 2016). Moreover, video IP cameras are installed at SMTJ, AHRJ, LLAJ, and JBAJ stations.

The RESAJ is the core of the Seismological Laboratory at the postgraduate program at SisVOc. As themes for Master's thesis, we are working to improve the algorithm to detect and obtain preliminary locations for earthquakes, and to use a

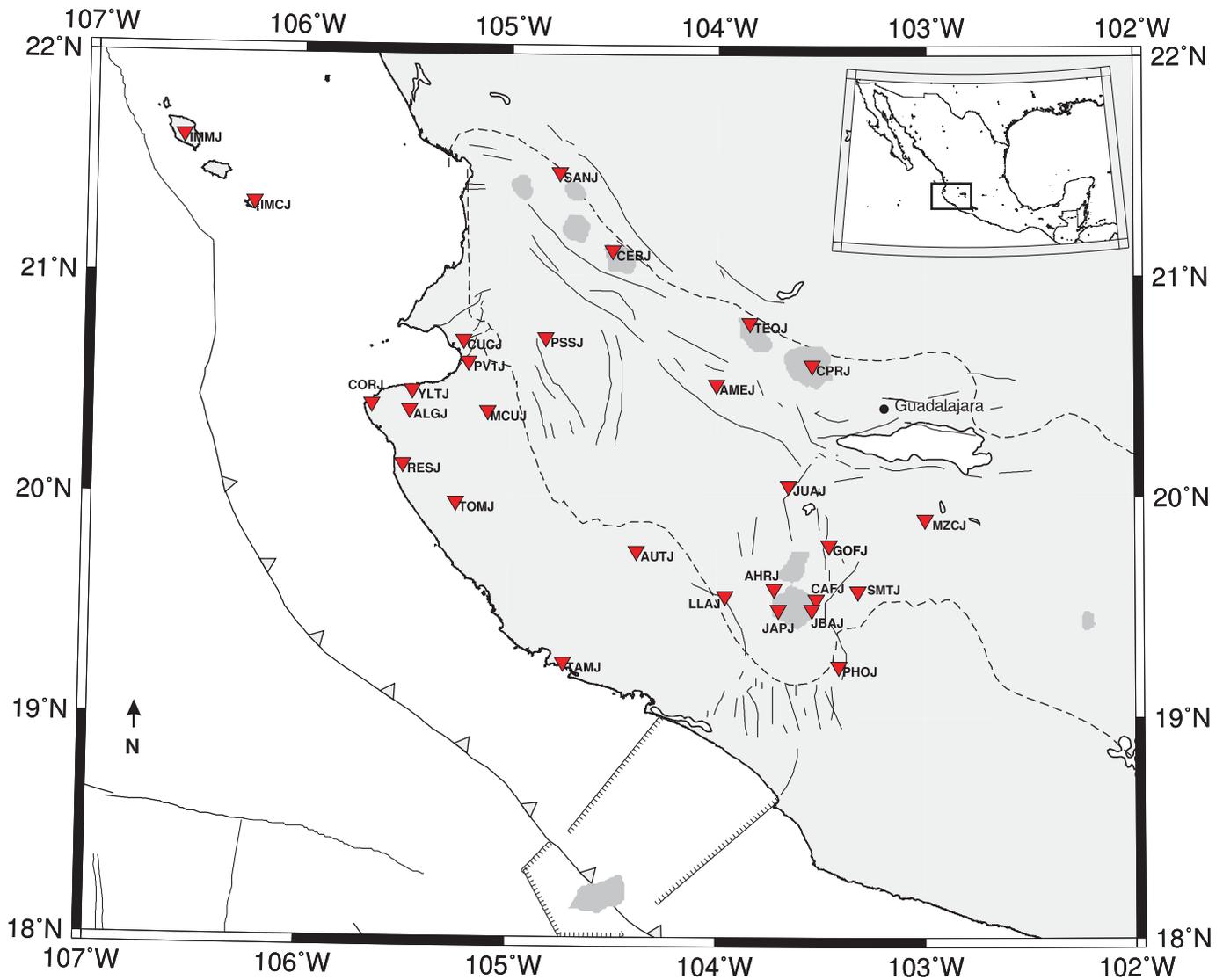

▲ **Figure 7.** Geographic distribution of the Red Sísmica y Acelerométrica Telemétrica de Jalisco (RESAJ) seismic stations.

velocity model determined from the TsuJal project (Núñez-Cornú et al., 2016). Additional Antelope applications in real time for seismic analysis are being adapted to address the specific characteristics of our network, and students are working on new applications to incorporate into the routine processing flow.

Seismicity studies for targeted sources using RESAJ data have been carried out as Master's theses (Ochoa-Chavez 2014; Tinoco Villa 2015; Córdoba Camargo, 2015; García-Millan, 2016; Acosta-Hernández, 2017; Rengifo, 2017; Marín Meza, 2017). A study of an earthquake swarm in the metropolitan area of Guadalajara (Zapopan) shows the seismic hazard associated with active tectonic structures not previously identified (Rengifo, 2017). Based on this work, an agreement with the municipality of Zapopan, Jalisco has been signed to install additional seismic stations and accelerometers there, and to train Zapopan Civil Defense staff for operations in a local observatory to be installed.

## DATA AND RESOURCES

All geophysical data collected by the Red Sísmica y Acelerométrica Telemétrica de Jalisco (RESAJ) are in a database at Centro de Sismología y Volcanología de Occidente (SisVOc) Posgraduate Program. The data may be available for use in collaborative research projects between SisVOc and other interested institutions by specific agreements. For information, contact pacornu77@gmail.com. The map data in Figure 6 were generated with Google Earth, whose data were provided by Lamont–Doherty Earth Observatory (LDEO)-Columbia, National Science Foundation (NSF), National Oceanic and Atmospheric Administration (NOAA), Scripps Institution

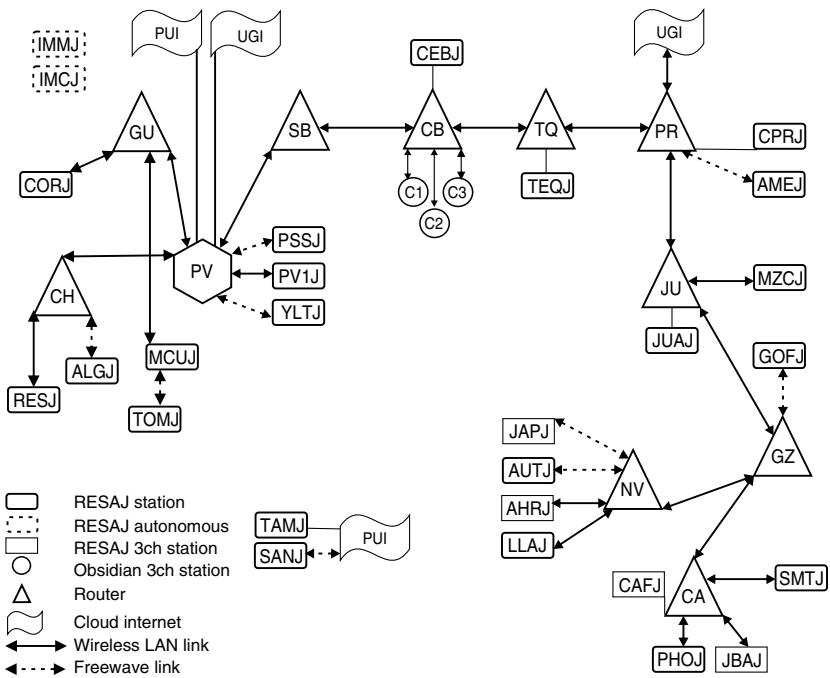

▲ Figure 8. Logistic diagram of RESAJ data transmission. PUI, Public internet network; UGI, University of Guadalajara internal internet network; CA, Cerro Alto; NV, Nevado; GZ, Cd. Guzman; JU, Juanacatlan; PR, Primavera; TQ, Tequila; CB, Ceboruco; SB, San Sebastian; GU, Guamuchil; CH, Chacala; PV, Puerto Vallarta.

of Oceanography (SIO), U.S. Navy, Next Generation Attenuation (NGA), and GEBCO.

## ACKNOWLEDGMENTS

The authors wish to thank Charlotte A. Rowe for their comments and suggestions, as well as anonymous reviewers for revisions that greatly improved the article.

The authors thank Quiriart Gutierrez, Juan Pinzon, and Yair Lopez for their assistance in the installation and maintenance of the stations.

This research is funded by Projects: CONACyT–FOMIXJal 2008-96567 (2009), CONACyT–FOMIXJal 2008-96539 (2009), CONACyT–FOMIXJal 2010-149245 (2010), and Universidad de Guadalajara (UdeG) internal projects; Edgar Alarcón was funded by a master scholarship from CONACYT, CVU 659870, Reg 421800. The authors thank H. Ayuntamiento de El Tuito, Jal., H. Ayuntamiento de Pihuamo, Jal., H. Ayuntamiento de Zapotitlán de Vadillo, Jal., H. Ayuntamiento de San Gabriel, Jal., H. Ayuntamiento de Puerto Vallarta, H. Ayuntamiento de San Sebastián del Oeste, Jal., and Unidad Municipal de Protección Civil de Tuxpan, Jal., for supporting to build the sheds. Unidad Estatal de Protección Civil y Bomberos de Nayarit y Unidad Estatal de Protección Civil y Bomberos de Jalisco for logistical support. The authors also acknowledge to Teléfonos de Mexico and C4 Gobierno de Nayarit for the facilities to transmit the data.

## REFERENCES

Acosta-Hernández, J. A. (2017). Estudio de la Corteza de la Zona Occidental de las Islas Marías Mediante Métodos Sísmicos, *Master Thesis*, Maestría en Ciencias en Geofísica, UdeG, 113 pp. (in Spanish).

Allan, J. F., J. Nelson, J. Luhr, J. Carmichael, M. Wopat, and P. Wallace (1991). Pliocene-recent rifting in SW México and associated volcanism: An exotic terrane in the making, *Am. Assoc. Petrol. Geol. Bull. Mem.* **47,** 425–445.

Bourgois, J., D. Renard, J. Auboin, W. Bandy, E. Barrier, T. Calmus, J. C. Carfantan, J. Guerrero, J. Mammerickx, B. Mercier de Lepinay, *et al.* (1988). Fragmentation en cours du bord Ouest du Continent Nord Americain: Les frontières sous-marines du Bloc Jalisco (Mexique), *C. R. Acad. Sci. Paris* **307,** no. II, 1121–1133 (in French).

Córdoba Camargo, A. A. (2015). Patrones Sísmicos en la Zona de Cabo Corrientes Jalisco, *Master Thesis*, Maestría en Ciencias en Geofísica, UdeG, 83 pp. (in Spanish).

Dañobeitia, J. J., R. Bartolomé, M. Prada, F. J. Nuñez-Cornú, D. Córdoba, W. L. Bandy, F. Estrada, A. L. Cameselle, D. Nuñez, A. Castellón, *et al.* (2016). Crustal architecture at the collision zone between Rivera and North American plates at the Jalisco block: Tsujal project, *Pure Appl. Geophys.* **173,** 3553–3573, doi: 10.1007/s00024-016-1388-7.

DeMets, C., and S. Stein (1990). Present-day kinematics of the Rivera plate and implications for tectonics of souhtwestern Mexico, *J. Geophys. Res.* **95,** 21,931–21,948.

García-Millan, N. (2016). Análisis de Ondas Coda en el Occidente de México, *Master Thesis*, Maestría en Ciencias en Geofísica, UdeG, 88 pp. (in Spanish).

Garduño, V. H., and A. Tibaldi (1991). Kinematic evolution of the continental active triple junction of the western Mexican volcanic belt, *C. R. Acad. Sci. Paris*, Série II, 135–142.

Gutierrez, Q. J., C. Escudero, and F. J. Núñez-Cornú (2015). Geometry of the Rivera-Cocos subduction zone inferred from local seismicity, *Bull. Seismol. Soc. Am.* **105,** no. 6, 3104–3113, doi: 10.1785/0120140358.

Kostoglodov, V., and W. Bandy (1995). Seismotectonic constraints on the convergence rate between the Rivera and North American plates, *J. Geophys. Res.* **100,** no. B9, 17,977–17,989.

Lennartz Electronic GmbH (2016). *LExD Seismometer Family. Document Number: 990-0073*, 30 pp.

Lindquist, K. G., R. L. Newman, and F. L. Vernon (2007). The antelope interface to PHP and applications: Web-based real-time monitoring, *Seismol. Res. Lett.* **78,** no. 6, 663–670, doi: 10.1785/gssrl.78.6.663.

Luhr, J., S. Nelson, J. Allan, and I. Carmichael (1985). Active rifting in southwestern Mexico: Manifestations of an incipient eastward spreading-ridge jump, *Geology* **13,** 54–57.

Marín Meza, T. (2017) Análisis Sísmico en el Bloque de Jalisco, período junio–diciembre 2015, *Master Thesis*, Maestría en Ciencias en Geofísica, UdeG, 70 pp. (in Spanish).

Mortera Gutiérrez, C. A., W. L. Bandy, F. Ponce Núñez, and D. A. Pérez Calderón (2016). Bahía de Banderas, Mexico: Morphology, magnetic anomalies and shallow structure, *Pure Appl. Geophys.* **173,** 3525–3551, doi: 10.1007/s00024-016-1384-y.

Núñez-Cornú, F. J. (2011). Peligro sísmico en el bloque de Jalisco, *Física de la Tierra* **23,** 199–229, doi: 10.5209/rev_FITE.2011.v23.36919 (in Spanish).

Núñez-Cornú, F. J. (2017). Jalisco, Mexico: A review of historical seismicity (M > 7.0), *Unpublished Manuscript*.

Núñez-Cornú, F. J., D. Córdoba, J. J. Dañobeitia, W. Bandy, M. Ortiz-Figueroa, R. Bartolomé, D. Núñez, A. Zamora-Camacho, J. M. Espíndola, A. Castellón, et al. (2016). Geophysical studies across Rivera plate and Jalisco block, Mexico: TsuJal Project, *Seismol. Res. Lett.* **87,** no. 1, 59–72, doi: 10.1785/0220150144.

Núñez-Cornú, F. J., R. L. Marta, F. A. Nava, G. Reyes-Dávila, and C. Suárez-Plascencia (2002). Characteristics of the seismicity in the coast and north of Jalisco block, Mexico, *Phys. Earth Planet. In.* **132,** 141–155.

Núñez-Cornú, F. J., R. M. Prol, A. Cupul-Magaña, and C. Suarez-Plascencia (2000). Near Shore submarine hydrothermal activity in Bahia Banderas, *Geofís. Int.* **39,** no. 2, 171–178.

Núñez-Cornú, F. J., G. A. Reyes-Dávila, M. Rutz López, E. Trejo Gómez, M. A. Camarena-García, and C. A. Ramírez-Vazquez (2004). The 2003 Armería, México earthquake ($M_w$ 7.4): Mainshock and early aftershocks, *Seismol. Res. Lett.* **75,** 734–743.

Núñez-Cornú, F. J., M. Rutz-López, V. Márquez-Ramírez, C. Suárez-Plascencia, and E. Trejo-Gómez (2010). Using an enhanced dataset for reassessing the source region of the 2003 Armería, Mexico earthquake, *Pure Appl. Geophys.* **168,** 1293–1302, doi: 10.1007/s00024-010-0178-x.

Núñez-Cornú, F. J., C. Suárez-Plascencia, C. R. Escudero Ayala, and A. Gómez (2011). Jalisco Regional Seismic Network (RESAJ), *Eos Trans. AGU* **92,** no. 52, Abstract S51A–2181.

Ochoa-Chavez, J. (2014). Tomografía de la Estructura de Velocidades de la Onda P en la Corteza Continental del Bloque Jalisco, *Master Thesis*, Maestría en Ciencias en Geofísica, UdeG, 59 pp. (in Spanish).

Ochoa-Chávez, J., C. Escudero, F. J. Núñez-Cornú, and W. Bandy (2016). *P*-wave velocity tomography from local earthquakes in western Mexico, *Pure Appl. Geophys.* **173,** 3487–3511, doi: 10.1007/s00024-015-1183-x.

PCJal (2016). *Sistema de Alertamiento de Tsunamis y Ciclones Tropicales*, available at http://proteccioncivil.jalisco.gob.mx/monitoreo/sistema-de-alertamiento-de-tsunamis.

Pinzón, J., F. J. Núñez-Cornú, and C. A. Rowe (2016). Magma intrusion near Volcan Tancítaro: Evidence from seismic analysis, *Phys. Earth Planet. In.* **262,** 66–79, doi: 10.1016/j.pepi.2016.11.004.

Reginfo Alcantara, W. M. (2017). Estudio de la Secuencia Sísmica Ocurrida en el Norte de la Zona Metropolitana de Guadalajara (ZMG) el 11 de Mayo de 2016, *Master Thesis*, Maestría en Ciencias en Geofísica, UdeG, 50 pp. (in Spanish).

Rutz López, M., and F. Núñez-Cornú (2004). Sismotectónica del norte y oeste del Bloque de Jalisco usando datos sísmicos regionales, *GEOS* **24,** no. 1, 2–13 (in Spanish).

Singh, S. K., L. Ponce, and S. P. Ninshenko (1985). The Great Jalisco, Mexico earthquakes of 1932: Subduction of the Rivera plate, *Bull. Seismol. Soc. Am.* **75,** 1301–1313.

Suter, M. (2015). The A.D. 1567 $M_w$ 7.2 Ameca, Jalisco, earthquake (western Trans-Mexican volcanic belt). Surface rupture parameters, seismological effects, and macroseismic intensities from historical sources, *Bull. Seismol. Soc. Am.* **104,** no. 2A, 646–656, doi: 10.1785/0120140163.

Tinoco Villa, M. E. (2015). Detección y Análisis de Enjambres Sísmicos en la Corteza Oceánica al Sur de las Islas Marías, *Master Thesis*, Maestría en Ciencias en Geofísica, UdeG, 69 pp. (in Spanish).

Trejo-Gómez, E., M. Ortiz, and F. J. Núñez-Cornú (2015). Source model of the October 9, 1995 Jalisco-Colima tsunami as constrained by field survey reports, and on the numerical simulation of the tsunami, *Geofís. Int.* **54,** no. 2, 153–162.

Urías Espinosa, J., W. L. Bandy, C. A. Mortera-Gutiérrez, Fco. J. Núñez-Cornú, and N. C. Mitchell (2016). Multibeam bathymetric survey of the Ipala Submarine Canyon, Jalisco, Mexico (20°N): The southern boundary of the Banderas forearc block, *Tectonophysics* **671,** 249–263, doi: 10.1016/j.tecto.2015.12.029.

White, R., and W. McCausland (2016). Volcano-tectonic earthquakes: A new tool for estimating intrusive volumes and forecasting eruptions, *J. Volcanol. Geoth. Res.* **309,** 139–155, doi: 10.1016/j.jvolgeores.2015.10.020.

*Francisco Javier Núñez-Cornú*
*Juan Manuel Sandoval*
*Edgar Alarcón*
*Adán Gómez*
*Carlos-Plascencia Suarez*
*Diana Núñez Escribano*
*Elizabeth Trejo-Gómez*
*Oscar Sánchez Mariscal*
*J. Guadalupe Candelas Ortiz*
*Luz María Zúñiga-Medina*
*Centro de Sismología y Volcanología de Occidente*
*SisVOc Universidad de Guadalajara*
*Puerto Vallarta*
*Jalisco, Mexico*
*pacornu77@gmail.com*

# QUERIES

1. AU: Please verify the name(s) of the authors as edited and correct if necessary.
2. AU: Please check the inserted affiliation and provide complete postal service address for the affiliation.
3. AU: Please indicate if the roman capital M throughout the article should be changed to (1) bold **M** or (2) *M*w (italic "M" and subscript roman "w").
4. AU: Please check meaning. Not sure if it means the rupture area was the entire coast, or if it was just located within the coast
5. AU: As per SSA style, the abbreviation "(VG)" has been deleted because it is not used again in this article.
6. AU: The spelling "Mortera-Gutierrez" was changed to match that in the reference ("Mortera-Gutiérrez"). Please let us know if the preferred spelling should be "Mortera-Gutierrez."
7. AU: The citation "Rutz and Núñez–Cornú (2004)" does not have a corresponding Reference entry. There is a "Rutz López and Núñez–Cornú (2004)" in the References, which is not cited in the paper. Please (1) decide whether these refer to the same work and (2) indicate the required changes to the paper and the References.
8. AU: As per SSA style, the abbreviation "(MARS)" has been deleted because it is not used again in this article.
9. AU: The spelling "Ochoa-Chavez" was changed to match that in the reference ("Ochoa-Chávez"). Please let us know if the preferred spelling should be "Ochoa-Chavez."
10. AU: The citation "Pinzon *et al.*, 2017" does not have a corresponding Reference entry. There is a "Pinzón *et al.*, 2016" in the References, which is not cited in the paper. Please (1) decide whether these refer to the same work and (2) indicate the required changes to the paper and the References.
11. AU: As per SSA style, the abbreviation "(CENAPRED)" has been deleted because it is not used again in this article.
12. AU: As per SSA style, the abbreviation "(FOPREDEN)" has been deleted because it is not used again in this article.
13. AU: As per SSA style, the abbreviation "(COECYTJal)" has been deleted because it is not used again in this article.
14. AU: As per SSA style, the abbreviation "GPS" has been deleted because it is not used again in this article.
15. AU: As per SSA style, the abbreviation "ISC" has been deleted because it is not used again in this article.
16. AU: As per SSA style, the abbreviation "STA/LTA" has been deleted because it is not used again in this article.
17. AU: Should this be in "all of the regions"? if more than one region
18. AU: The spelling "Cordoba Camargo" was changed to match that in the reference ("Córdoba Camargo"). Please let us know if the preferred spelling should be "Cordoba Camargo."
19. AU: The spelling "Acosta-Hernandez" was changed to match that in the reference ("Acosta-Hernández"). Please let us know if the preferred spelling should be "Acosta-Hernandez."
20. AU: The citation "Rengifo (2017)" does not have a corresponding Reference entry. There is a "Reginfo Alcantara (2017) " in the References, which is not cited in the paper. Please (1) decide whether these refer to the same work and (2) indicate the required changes to the paper and the References.
21. AU: The spelling "Marin Meza" was changed to match that in the reference ("Marín Meza"). Please let us know if the preferred spelling should be "Marin Meza."
22. AU: Please provide a definition of "GEBCO"; it will be included before the abbreviation.
23. AU: For Núñez-Cornú (2017): SSA cannot include unpublished manuscripts in the References section. If the article has not been accepted for publication by the time the article reaches the final review stage, then the article will need to be cited as an unpublished manuscript and details included in the Data and Resources section (if desired). Citations would be changed to a format such as "... from F. J. Núñez-Cornú, unpublished manuscript, 2017; see Data and Resources)."
24. AU: For PCJal (2016): Please provide the month and year when you last accessed this website for this reference in your article.
25. AU: Please clarify whether the edits to Table 1 do not affect your intended meaning.
26. AU: It is SSA style to set dates as (yyyy/mm/dd). Column 2 has been reformatted accordingly. Please review the changes to be sure all were made correctly.
27. AU: Although Singh et al (1984) is cited, there is no corresponding Reference entry. Please provide a Reference entry for this citation, or indicate the citations should be deleted throughout the article.